\documentclass[namedreferences]{solarphysics}
\usepackage[hyperref,optionalrh,solaromanenum]{spr-sola-addons} % For Solar Physics
\usepackage{graphicx}                    % For eps figures, newer & more powerfull
\usepackage{color}                       % For color text: \color command
\usepackage{breakurl}                         % For breaking URLs easily trough lines
\begin{document}
\sloppy
\begin{article}

\begin{opening}

\title{Non-axisymmetric component of the photospheric magnetic field}

%%%%%%%%%%%%%%%%%%%%%%%%%%%%%%%%%%%%%%%%%%%%%%%%%%%
%% Authors Names
%
\author[addressref={aff1},corref,email={helena@ev13934.spb.edu}]{\inits{E.S.}\fnm{E.S.}\lnm{Vernova}\orcid{0000-0001-8075-1522}}
\author[addressref=aff1]{\inits{M.I.}\fnm{M.I.}~\lnm{Tyasto}}\sep
\author[addressref=aff2]{\inits{D.G.}\fnm{D.G.}~\lnm{Baranov}\orcid{0000-0003-2838-8513}}
\author[addressref=aff1]{\inits{O.A.}\fnm{O.A.}~\lnm{Danilova}}\sep

%%%%%%%%%%%%%%%%%%%%%%%%%%%%%%%%%%%%%%%%%%%%%%%%%%%
%% Runningheads
%
\runningauthor{E.S.~Vernova \textit{et al.}} \runningtitle{Non-axisymmetric component of the photospheric magnetic field}

%%%%%%%%%%%%%%%%%%%%%%%%%%%%%%%%%%%%%%%%%%%%%%%%%%%
%% Affilations
%% id shold be the same with \author addressref value.
\address[id={aff1}]{IZMIRAN, SPb. Filial, Laboratory of Magnetospheric
Disturbances, St. Petersburg, Russia}
\address[id={aff2}]{Ioffe Institute, St. Petersburg, Russia}
%%%%%%%%%%%%%%%%%%%%%%%%%%%%%%%%%%%%%%%%%%%%%%%%%%%
%%% Abstract
\begin{abstract}
The longitudinal asymmetry of the photospheric magnetic field distribution is studied on the basis of the data of the Kitt Peak National Solar
Observatory (synoptic maps for the period 1976--2016). The method of vector summing of magnetic fields is used, which allows to decrease the
influence of the stochastic components uniformly distributed over the whole longitude interval, and to stress a steady non-axisymmetric
component of the field. The distributions of magnetic fields of different intensity are considered separately for the strong ($B > 50$\,G), weak
($B < 5$\,G) and medium ($50 > B > 5$\,G) fields. It is shown that the longitudinal asymmetry for all groups of fields changes in phase with the
11-year cycle of solar activity. The asymmetry of strong and medium fields changes in phase with magnetic fluxes of these fields, while the
asymmetry of weak fields is in antiphase with the flux of weak fields. The distributions of strong and medium magnetic fields over the longitude
are similar in the shape: The maximum of distribution is located at the longitude $\sim 180^\circ$ during the period of ascent--maximum of solar
cycle and at the longitude $\sim 0^\circ/360^\circ$ during the period of descent--minimum. The weak fields exhibit the opposite picture: the
maximum of their distribution is always observed at the longitude, where the strong and medium fields show minimum.
\end{abstract}

%%%%%%%%%%%%%%%%%%%%%%%%%%%%%%%%%%%%%%%%%%%%%%%%%%%
%% Keywords
%

\end{opening}
%-------------------------------------------------

%%%%%%%%%%%%%%%%%%%%%%%%%%%%%%%%%%%%%%%%%%%%%%%%%%%
%% Sections
%
\section{INTRODUCTION}\label{intro}
The models of solar dynamo based on the assumption of the axisymmetric magnetic field of the Sun successfully describe main features of solar
activity  evolution: the genesis of 11-year and 22-year cycles and a number of phenomena connected with these cycles (see the review by
\inlinecite{char10}). During 11-year cycle not only intensity of different manifestations of solar activity changes, but also their distribution
over the Sun's surface. For example, the width of the sunspot zone and its mean latitudinal location change in the course of a solar cycle.
However models of the axisymmetric dynamo do not explain the appearance of longitudinal asymmetry which was observed in distributions of
sunspots, solar flares and other manifestations of solar activity as well as in the distribution of the solar magnetic fields. It was shown that
the non-axisymmetric mode $m=1$ is always present in the solar photospheric magnetic field and during a solar maximum this mode dominates in
comparison with the axisymmetric mode $m=0$ (\opencite{ruz01}).

Non-axisymmetry of solar activity appears, in particular, in the form of so-called active longitudes (preferred longitudes). Active longitudes
usually imply rather narrow range of solar longitudes ($20^\circ$--$60^\circ$), where the different manifestations of solar activity occur more
often, than at other longitudes (\opencite{vit69}; \opencite{ben02a}, and references therein). The lifetime of active longitudes considerably
varied in different studies -- from several years to several solar cycles (see, e.g., \opencite{bum65}; \opencite{bai03}).

The tendency of solar activity to cluster around longitudes separated by $180^\circ$ was found by several authors (\opencite{jet97};
\opencite{mor01}; \opencite{bai03}). The long-lived complexes of solar activity were studied by \inlinecite{ben99}, \inlinecite{ben13}. These
complexes connected with magnetic fields located under the photosphere were observed on the same longitudes both at the end of a solar cycle
(near the minimum of activity) and at the beginning  of the new cycle.

It was shown that the active longitudes are rotating approximately with the speed of the Carrington rotation (\opencite{gai83};
\opencite{pip15}). On the contrary, \inlinecite{ber03} came to the conclusion that for latitudes which are different from the mean latitude of
the sunspot zone the active longitudes, separated by $180^\circ$, migrate relative to the Carrington coordinate system due to the differential
rotation of the Sun and the change of rotation speed.

There is a hypothesis that the longitudinal asymmetry forms due to the influence of the inclined relict magnetic field of the Sun. This field
could be a cause of north-south asymmetry, appearance of active longitudes, and 22-year cycle of solar activity (\opencite{kit01};
\opencite{mur01}).

Intensive search for the model of the non-axisymmetric dynamo  is now in progress (\opencite{big04}; \opencite{jia07}; \opencite{pip15}), the
problem of active longitudes on the Sun being considered as one of the important points in building of such a model. The active longitudes are
named to be one of the most interesting manifestations of the non-axisymmetric magnetic field of the Sun (\opencite{pip15}). It is possible to
make a conclusion that the modern model of the solar dynamo necessarily should explain such phenomena, as origin of active longitudes and their
periodic displacement by $180^\circ$ ("flip-flop" effect).

This article is a continuation and extension of our previous studies of the asymmetry of solar activity and magnetic field (\opencite{ver02},
\opencite{ver07}). The non-axisymmetric component of the photospheric magnetic field is investigated using the method of vector summing of
magnetic fields. This method allows to decrease the contribution of stochastic components of the magnetic field and to underline a role of the
non-axisymmetric field components. The time evolution of the longitudinal asymmetry during four solar cycles and localization of active
longitudes are studied.

\section{DATA AND METHOD}

The longitudinal asymmetry of   magnetic field distribution over the photosphere was considered on the basis of synoptic maps which are produced
by the Kitt Peak National Solar Observatory (NSO Kitt Peak). Combination of the data obtained by two devices of NSO Kitt Peak -- KPVT (Kitt Peak
Vacuum Telescope) for 1976--2003  (ftp://nispdata.nso.edu/kpvt/synoptic/mag/) and SOLIS (Synoptic Optical Long-term Investigations of the Sun)
for 2003--2016  (https://magmap.nso.edu/solis/archive.html) allowed to study the longitudinal asymmetry of the magnetic fields for four solar
cycles. Each map contained 180 intervals of sine latitude uniformly distributed between $+90^\circ$ and $-90^\circ$. The resolution of synoptic
maps in longitude was $1^\circ$. Thus, each map consisted of $360\times180$ pixels of the magnetic field in gauss.

For the evaluation of longitudinal asymmetry of  the  magnetic field distribution it is important to reduce the contribution of random
(stochastic) components, whose distribution over the longitude is essentially uniform. Vector summation of magnetic fields, presented as
synoptic maps, allows to stress a steady, regular part of asymmetry (\opencite{ver02}; \opencite{ver07}). We used this method to study the
non-axisymmetric components of the magnetic field. Synoptic maps were averaged over the latitude, giving the mean value of the magnetic field
$B_i$  ($i=1:360$) for $1^\circ$  intervals of longitude (longitudinal resolution of synoptic maps). The vector $\vec{B}_i$ was associated with
each of this intervals. The modulus of the vector was set equal to $B_i$ and the phase angle was defined as the  longitude of the interval. The
vector of longitudinal asymmetry $\vec{LA}$ (Longitudinal Asymmetry) was evaluated for each Carrington rotation of the Sun as the vector sum of
elementary vectors $\vec{B}_i$:
\begin{equation}
\label{1} \vec{LA} = \sum_{i=1}^{360} \vec{B}_i
\end{equation}
It is of interest to consider both characteristics of the longitudinal asymmetry -- the modulus of the vector and the phase.

Variations over time of the modulus of the vector $\vec{LA}$ allow to analyze the connection of non-axisymmetric components of the magnetic
field with solar cycle. The phase of  the vector points to the dominating longitude in a given Carrington rotation. The distribution of the
vector phase over longitude reveals the regions of high concentration of the magnetic fields (active longitudes). When analyzing the
longitudinal distribution of the magnetic fields, the values of  the vector modulus  below a given threshold were rejected. The threshold was
selected so that the number of the rejected cases made $\sim 15$ \%. The low values of asymmetry can be observed in two cases: 1. General
decrease of the strength of the photospheric magnetic field during magnetic cycle of the Sun; 2. The presence of several regions of the magnetic
field concentration located on a longitude in such a way that their contributions are mutually cancelled. In both cases the orientation of a
resultant vector will be subject to the influence of random factors and the location of an active longitude can not be determined with
reliability.

\section{RESULTS AND DISCUSSION}

\begin{figure}[t]
\begin{center}
\includegraphics[width=0.95\textwidth]{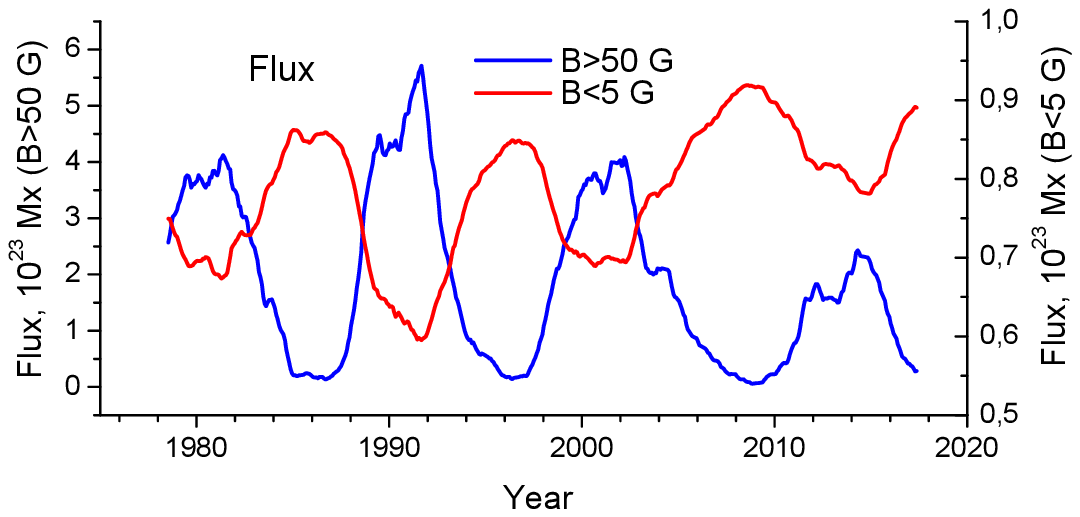}
\caption{Photospheric magnetic fields in 1976--2016: magnetic fluxes of the strong ($B > 50$\,G -- the blue line), and the weak ($B < 5$\,G --
the red line) fields.} \label{flux}
\end{center}
\end{figure}

In our article (\opencite{ver18}) the temporal development of magnetic fluxes for different groups of magnetic fields differing in strength was
studied. It is of interest to compare time changes of the magnetic flux and longitudinal asymmetry for the magnetic fields of different strength
B. In Figure~\ref{flux} the fluxes of the strong ($B > 50$\,G) and weak ($B < 5$\,G) magnetic fields are shown (the blue line corresponds to the
strong fields, and the red line -- to the weak fields). The strong magnetic fields change in  phase with solar activity cycle, reaching a
maximum around the second Gnevyshev maximum, following the so-called Gnevyshev gap. In contrast, the flux of weak magnetic fields changes in
antiphase with solar activity cycle and with the flux of strong fields (coefficient of correlation between fluxes $R = -0.91$). Similar
conclusions were made by \inlinecite{jin14}, who show that the magnetic structures with low fluxes change in antiphase with solar cycle. We do
not present here the time course of the magnetic fields with a medium  strength $50 > B > 5$\,G, as it repeats changes of the strong magnetic
fields  being also in  antiphase with the weak fields (coefficient of correlation between fluxes of the medium and weak fields $R =-0.98$).
%%%%%%%%%%%%%%%%%%%%%%%%%%%%%%%%%%%%%%%%%% FIGURE 1
%%%%%%%%%%%%%%%%%%%%%%%%%%%%%%%%%%%%%%%%%%%%%%%%%%%%%%
%%%%%%%%%%%%%%%%%%%%%%%%%%%%%%%%%%%%%%%%%% FIGURE 2
\begin{figure}[t]
\begin{center}
\includegraphics[width=0.95\textwidth]{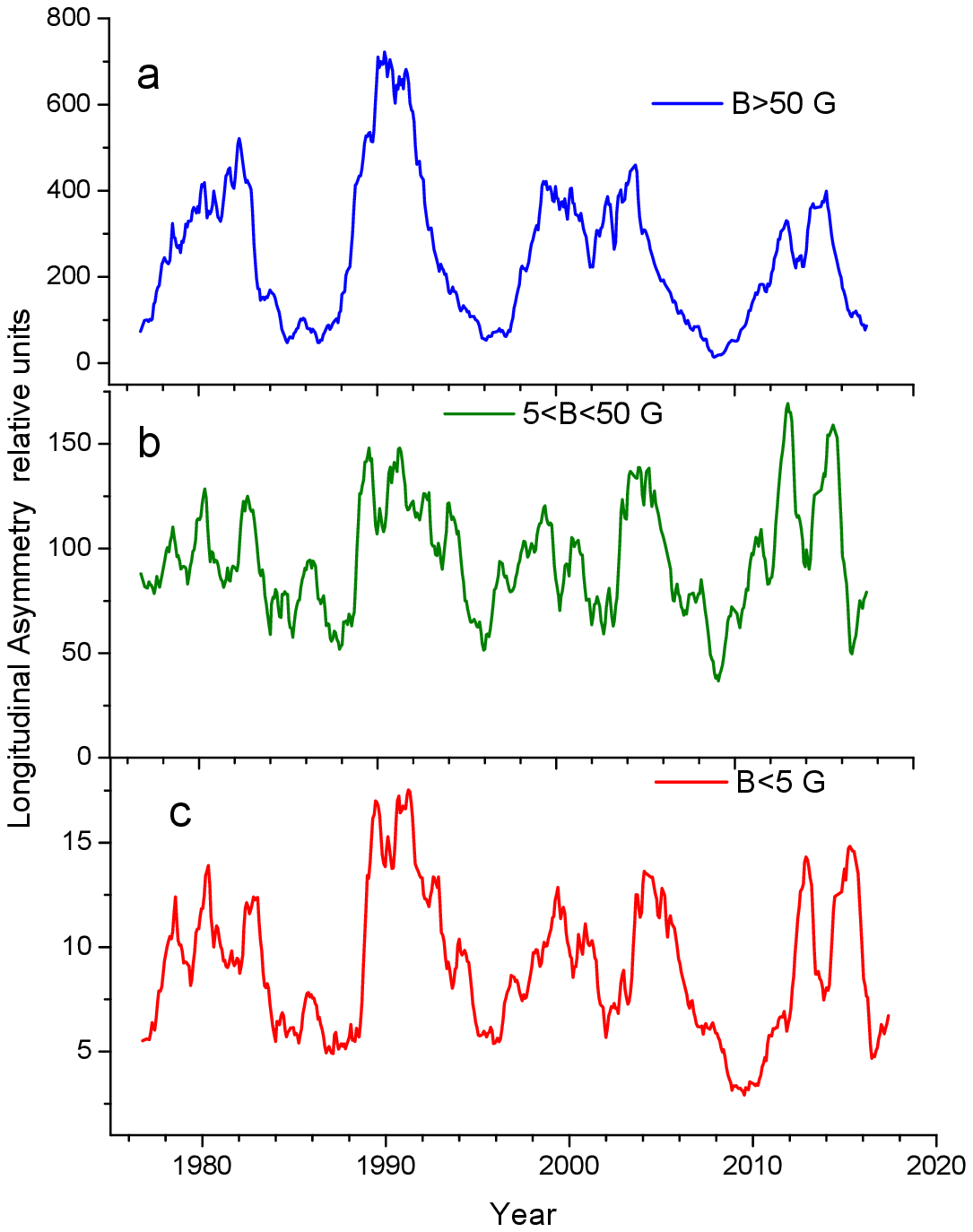}
\caption{Longitudinal asymmetry $\vec{LA}$ of the magnetic fields a) for the strong ($B > 50$\,G), b) medium ($50 > B > 5$\,G) and c) weak ($B <
5$\,G) magnetic fields in Solar Cycles 21--24.} \label{groups}
\end{center}
\end{figure}
%%%%%%%%%%%%%%%%%%%%%%%%%%%%%%%%%%%%%%%%%%%%%%%%%%%%%%
The development  of the longitudinal asymmetry is presented in Figure~\ref{groups} for three groups of fields: for the strong fields ($B >
50$\,G), for the weak fields ($B < 5$\,G), and also for the fields of medium strength ($50 > B > 5$\,G). Contrary to the magnetic fluxes
(Figure~\ref{flux}) the longitudinal asymmetries for the strong, medium and weak fields change synchronously. For all three groups of fields the
longitudinal asymmetry (Figure~\ref{flux}a, b, c), as well as the flux of the strong fields (Figure~\ref{flux} the blue line), change with
11-year cycle of solar activity. This result agrees with the model of kinematic generation (dynamo) offered by \inlinecite{big04}. In this
article it is indicated that the non-axisymmetric component of the magnetic field plays an important role in the origin of solar activity. In
particular, the non-axisymmetric mode follows a course of solar cycle as a result of  coupling between non-axisymmetric and symmetric modes.

Unexpectedly, the longitudinal asymmetry of the weak fields (Figure~\ref{groups}c)  follows not the flux changes of the weak fields
(Figure~\ref{flux}, the red line), but the general course of solar cycle and correlates well with the longitudinal asymmetry of strong fields
($R=0,86$).

When compared with the change over time of the magnetic flux, the change of the longitudinal asymmetry during a solar cycle exhibits a number of
specific features. Even after the same running averaging over 13 rotations the time course of the longitudinal asymmetry shows large
fluctuations, especially for the magnetic fields of medium strength, in contrast to the magnetic flux. Another feature of the longitudinal
asymmetry is a pronounced two-maxima structure of its time course. It is especially clearly seen in Cycle 23, where the two maxima are
substantially separated in time.

The study of time variations of the longitudinal asymmetry allowed to establish the following patterns:

1. The longitudinal asymmetry for all groups of magnetic fields ($B > 50$\,G, $50 > B > 5$\,G and $B < 5$\,G follows  the 11-year solar cycle
with a sharp fall between two maxima (Gnevyshev gap).

2. The longitudinal asymmetry of the weak fields $B < 5$\,G develops in antiphase with the magnetic flux of weak fields.

So far we considered the modulus of the vector of longitudinal asymmetry $|\vec{LA}|$ and its change in time. We now turn to the phase of the
vector, which indicates the dominant longitude in this Carrington rotation. To analyze the phase of the vector, we used histograms, showing the
number of cases (rotations) that fall in each longitude interval. The total longitudinal distribution of the phase over several solar cycles
does not show the presence of preferred longitudes: the distribution over longitude is almost uniform. However, the situation changes
significantly when one independently considers different phases of the solar cycle.

It was earlier shown for three solar cycles that the localization of active longitudes of magnetic fields depends on a phase of solar cycle
(\opencite{ver07}). In longitudinal distribution of the photospheric magnetic fields two specific periods appear: (a) the phase of
ascent--maximum (AM) and (b) the phase of descent--minimum (DM). In the present article the analysis of longitudinal distribution was extended
to the period 2003--2016. The obtained results allowed to increase reliability of estimations of longitudinal asymmetry and confirmed the
previously made conclusions.

The combination of the ascent phase with the maximum (AM) and of the descent phase with the minimum (DM) corresponds to a particular periodicity
in the change of polarity of magnetic fields of the Sun.

%%%%%%%%%%%%%%%%%%%%%%%%%%%%%%%%%%%%%%%%%% FIGURE 3
\begin{figure}[t]
\begin{center}
\includegraphics[width=0.95\textwidth]{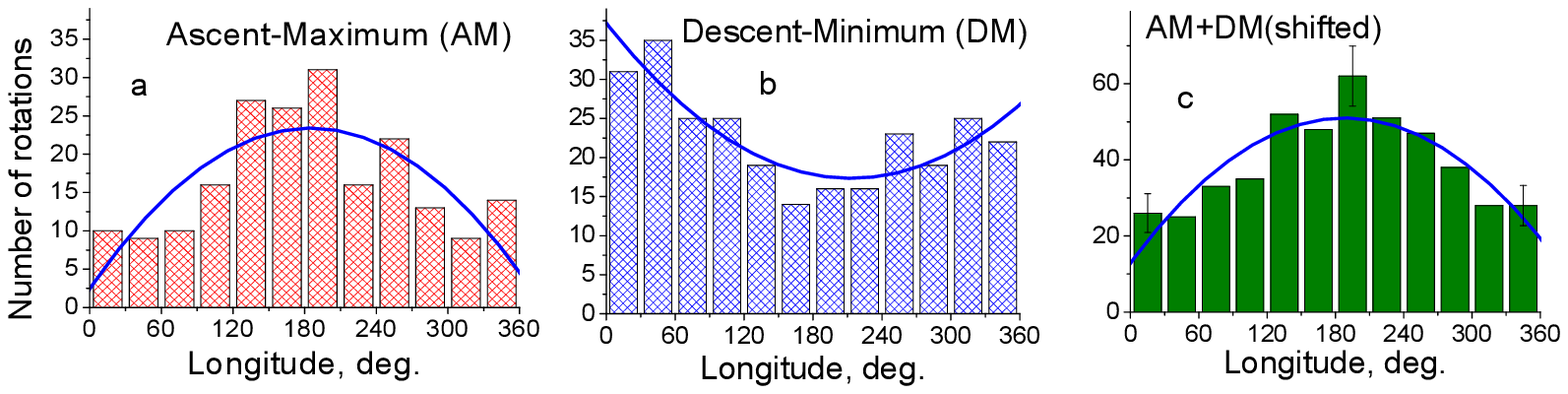}
\caption{Strong fields ($B > 50$\,G): longitudinal distribution of the photospheric magnetic field in 1976-2016 for the latitude interval $\pm
40^\circ$; a) phase of ascent and maximum of solar activity, AM; b) phase of descent and minimum, DM; c) summary total distribution for all
phases of solar cycle (summing of the histogram AM with the histogram DM shifted by $180^\circ$). Envelope curve -- approximation by a
polynomial of the second degree.} \label{strong}
\end{center}
\end{figure}
%%%%%%%%%%%%%%%%%%%%%%%%%%%%%%%%%%%%%%%%%%%%%%%%%%%%%%
%%%%%%%%%%%%%%%%%%%%%%%%%%%%%%%%%%%%%%%%%% FIGURE 4
\begin{figure}[t]
\begin{center}
\includegraphics[width=0.95\textwidth]{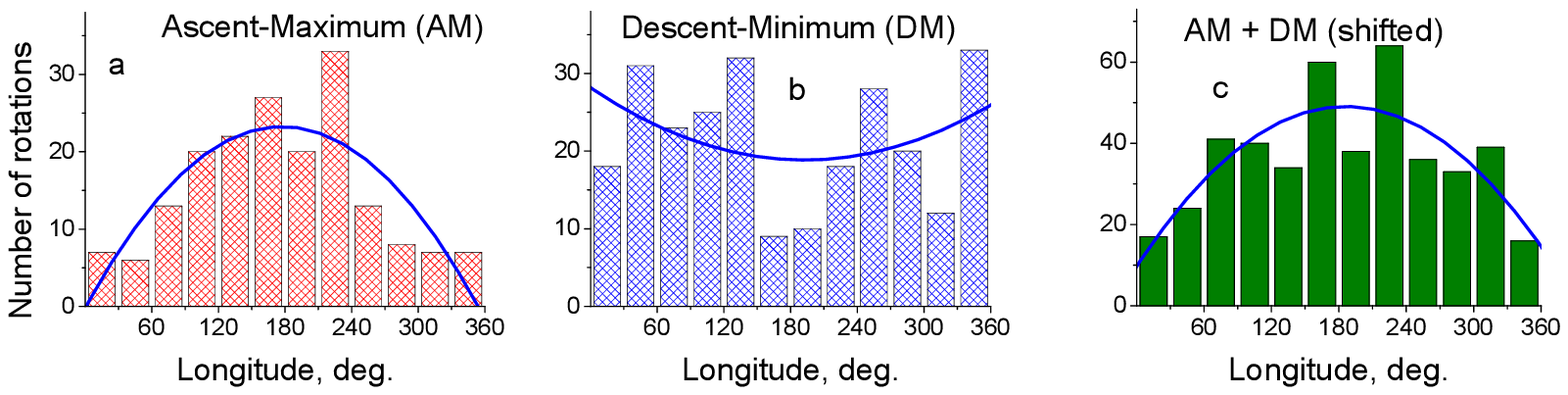}
\caption{Fields of medium intensity ($50 > B > 5$\,G): longitudinal distribution of the photospheric magnetic field in 1976--2016 for the
latitude interval $\pm 90^\circ$. a) phase of ascent and maximum of solar activity, AM; b) phase of descent and minimum, DM; c) summary
distribution for all phases of solar cycle (summing of the histogram AM with the histogram DM shifted by $180^\circ$). Envelope curve --
approximation by a polynomial of the second degree.} \label{med}
\end{center}
\end{figure}
%%%%%%%%%%%%%%%%%%%%%%%%%%%%%%%%%%%%%%%%%%%%%%%%%%%%%%
%%%%%%%%%%%%%%%%%%%%%%%%%%%%%%%%%%%%%%%%%% FIGURE 5
\begin{figure}[h!]
\begin{center}
\includegraphics[width=0.95\textwidth]{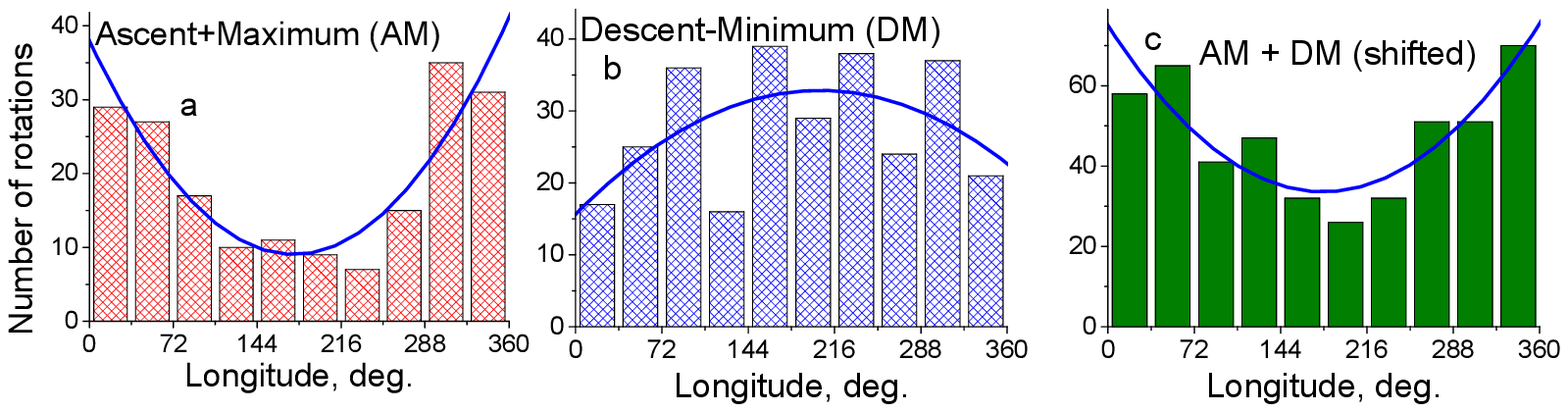}
\caption{Weak fields ($B < 5$\,G): longitudinal distribution of the photospheric magnetic field in 1976-2016 for the latitude interval $\pm
90^\circ$. a) phase of   ascent and maximum of solar activity, AM; b) phase of descent and minimum, DM; c) summary total distribution for all
phases of solar cycle (summing of the histogram AM with the histogram DM shifted by $180^\circ$). Envelope curve -- approximation by a
polynomial of the second degree.} \label{weak}
\end{center}
\end{figure}
%%%%%%%%%%%%%%%%%%%%%%%%%%%%%%%%%%%%%%%%%%%%%%%%%%%%%%
The time moments dividing the two specific intervals (ascent-maximum and descent-minimum) are the important critical points of the magnetic
solar cycle.They coincide with the change of the polar field sign near the maximum of solar cycle and with the change of signs of the leading
sunspots at the minimum of solar cycle (according to the Hale law).

The two specific periods -- AM and DM -- correspond to two different situations which are present in a 22-year magnetic cycle of the Sun, during
which the signs of the polar magnetic field and of the leading sunspots can be the same (in the given hemisphere) or opposite. During the period
from a minimum to an reversal the sign of the polar magnetic field and the sign of the magnetic field of the leading sunspots coincide, while
for the period from a reversal to a minimum the signs of these fields are opposite.

We considered distribution of the phase of the asymmetry vector over the longitude (Figures~\ref{strong},~\ref{med},~\ref{weak}) for three
groups of fields: for the strong fields ($B > 50$\,G, Figure~\ref{strong}), for the fields of medium strength ($50 > B > 5$\,G,
Figure~\ref{med}) and for the weak fields ($B < 5$\,G, Figure~\ref{weak}) (see time changes of longitudinal asymmetry of these groups of fields
in Figure~\ref{groups}a,b,c).

As the strong fields are connected with sunspots and are localized in the sunspot zone, for fields $'
> 50$\,G we confine ourselves with the latitude interval from
$-40^\circ$ to $+40^\circ$. The fields with smaller strength have no such well-defined localization in latitude, therefore for them all range of
latitudes from $-90^\circ$ to $+90^\circ$ was taken into account.

In Figure~\ref{strong}, the histograms of the phase of the longitudinal asymmetry for strong magnetic fields ($B > 50$\,G) are presented. For
the periods AM (ascent--maximum) and DM (descent--minimum) the maxima of distributions are located at two opposite Carrington longitudes:
$180^\circ$ (AM, Figure~\ref{strong}a) and $0^\circ/360^\circ$ (DM, Figure~\ref{strong}b). Shifting the DM histogram (Figure~\ref{strong}b) by
$180^\circ$ and summing it with the AM histogram (Figure~\ref{strong}a) we obtain the summary distribution of magnetic fields for both periods
of the solar cycle for 1976--2016 (Figure~\ref{strong}c). Clearly seen maximum in Figure~\ref{strong}c confirms displacement of the active
longitude by $180^\circ$. The stability of the location of active longitudes during four solar cycles can be considered as an evidence for their
rigid rotation with period approximately equal to the Carrington period.

In Figure~\ref{med}, the histograms for medium magnetic fields ($50 > B > 5$\,G) are presented. As well as in case of strong fields, the maxima
of distributions are located near two opposite Carrington longitudes: $180^\circ$ (AM, Figure~\ref{med}a) and $0^\circ/360^\circ$ (DM,
Figure~\ref{med}b). A comparison of the histograms for strong and medium magnetic fields shows their good agreement for the ascent--maximum
phase. For the descent--minimum phase the medium fields show a more complex distribution than the strong fields, however the form of the
envelope curves remains similar. Shifting the histogram DM (Figure~\ref{med}b) by $180^\circ$ and summing it with the histogram AM
(Figure~\ref{med}a) we obtain the summary distribution of magnetic fields for both periods of solar cycle for 1976--2016 (Figure~\ref{med}c).

The opposite picture is observed for the longitudinal distribution of the weak fields ($B < 5$\,G, Figure~\ref{weak}). As for the strong fields,
there is a shift of a maximum of distribution by $180^\circ$ when we pass from the phase AM (Figure~\ref{weak}a) to the phase DM
(Figure~\ref{weak}b). However, the maxima of the weak fields are located on the longitudes diametrically opposite in comparison with the strong
fields: at the longitude $0^\circ/360^\circ$ for the phase AM (Figure~\ref{weak}a) and at the longitude $180^\circ$ for the phase DM
(Figure~\ref{weak}b).

Shifting the histogram DM (Figure~\ref{weak}b) by $180^\circ$ and summing it with the histogram AM (Figure~\ref{weak}a) we obtain the summary
distribution of the magnetic fields for both periods of solar cycle for 1976-2016 (Figure~\ref{weak}c). Summary distributions of the weak fields
(Figure~\ref{weak}c) and the distributions of fields of greater strength (Figure~\ref{strong}c, Figure~\ref{med}c) have the opposite forms of
the envelopes.

The maximum concentration of fields is observed in turn at one of two opposite longitudes ($0^\circ/360^\circ$ and $180^\circ$). The change of
active longitudes by $180^\circ$ (flip-flop effect) happens at the time of the sign change of the leading and the following sunspots at the
minimum of solar activity and at the moment of the sign change of the  polar magnetic field. It is possible to assume that flip-flop of active
longitudes is directly connected to magnetic solar cycle.

The obtained results allow to connect the appearance of active longitudes with the solar cycle phases. For the strong fields, the main
conclusion of the work is that the longitudinal distribution of the photospheric magnetic field exhibits a maximum at the longitude $\sim
180^\circ$ for the ascent--maximum phase and at the longitude $\sim 0^\circ/360^\circ$  for the descent - minimum phase.

If we use for the phase of solar cycle the following notation: $k = 1$ corresponds to the period of 11-year cycle from the minimum to the polar
magnetic field reversal; $k = 2$ corresponds to the period from the reversal to the minimum, then the location of the distribution maximum (the
active longitude) can be represented by the formula:
\begin{equation}
\label{2}
 Active\,\, Longitude = \pi k
\end{equation}
The obtained relation agrees with the results of articles (\opencite{bai03}; \opencite{jet97}) where concentration of fields at longitudes
$180^\circ$ and $0^\circ/360^\circ$ was also observed. It is necessary to mention the article of \inlinecite{ski05} where on the basis of the
data on coronal mass ejections (CME) the increased activity of heliolatitudes $180^\circ$ and $30^\circ$ was detected. Moreover, these almost
diametrically opposite longitudes dominated in turn. Similarly to our results, the prevalence of the longitude $180^\circ$ coincided with the AM
phase of solar cycle, whereas the prevalence of the longitude $30^\circ$ coincided with the DM phase. The connection of the two specific periods
with changes of polarity of the global  magnetic field of the Sun and with the change of polarity of the leading sunspots hardly can be ascribed
to a mere coincidence.

\inlinecite{ben02b} showed that the relation between polarity of the global  magnetic field of the Sun and the leading sunspots polarity is of
fundamental importance. Using the data on soft x-radiation obtained by the Yohkoh x-ray telescope in 1991--2001, the authors of the article
found large closed magnetic loops connecting the magnetic flux of the following sunspots of an active region to the magnetic flux of the polar
region of the opposite polarity. These large scale magnetic connections appeared mainly during ascent phase of solar cycle and at its maximum.
During the descent phase of solar cycle these connections did not appear or were insignificant. Thus, two specific periods AM and DM correspond
to quite different patterns of the global  and local magnetic fields of the Sun.

\section{CONCLUSIONS}

1. The change of longitudinal asymmetry of the photospheric magnetic field for four solar cycles is considered. For the evaluation of the
longitudinal asymmetry the method of vector summing of fields was used. A vector was associated with every Carrington rotation whose modulus was
determined by the magnitude of asymmetry and whose phase indicated the position of the dominating longitude. This method allows to study the
non-axisymmetric component of the magnetic field, decreasing the influence of the stochastic components, uniformly distributed over the
longitude.

2. It is shown that the longitudinal asymmetry for the magnetic fields of different strength: strong ($B > 50$\,G), weak ($B < 5$\,G) and medium
($50 > B > 5$\,G) change in phase with the 11-year cycle of solar activity. The asymmetry of strong and medium fields changes in phase with the
magnetic fluxes of these fields, while the asymmetry of weak fields is in antiphase with the flux of weak fields.

3. The longitudinal distributions of the strong, medium and weak magnetic fields are studied. The strong and medium magnetic fields display
distributions similar in the form  which show a maximum at the longitude $\sim 180^\circ$ for the period ascent--maximum of the solar cycle and
at the longitude $\sim 0^\circ/360^\circ$ for the period descent--minimum. The weak fields show the opposite picture: the maximum of their
distribution is always located at the longitude of strong and medium fields minimum.

4. The longitude of a distribution maximum of the magnetic fields changes by $180^\circ$ two times during solar cycle: during the minimum of
solar activity and near the reversal of the polar field.

5. The obtained results reveal regular changes of longitudinal asymmetry of magnetic fields and close connection of these changes with the solar
cycle.
\bigskip
\\
{\bf Acknowledgements} NSO/Kitt Peak data used here are produced cooperatively by NSF/NOAO, NASA/GSFC, and NOAA/SEL. This work makes also use of
SOLIS data obtained by the NSO Integrated Synoptic Program (NISP), managed by the National Solar Observatory.

Disclosure of Potential Conflicts of Interest The authors declare that they have no conflicts of interest.

\end{article}
\end{document}